\documentclass[aip,reprint]{revtex4-1}
\draft 
 \usepackage{amsmath}
\begin{document}
\title{Direct evaluation of overlap integrals between Slater-type-orbitals} 
\author{Michael J. Caola}
\email[]{caolam@blueyonder.co.uk}
\affiliation{6, Normanton Rd, Bristol BS8 2TY, U.K.}
\date{\today}
\begin{abstract}
We derive direct single-stage numerical evaluation of the electronic overlap integral between arbitrary atomic orbitals (including STOs). Integration is over cartesian co-ordinates, and replaces previous sums over 'special' functions. The results, in Mathematica 10 and Maple 18, agree with the literature to $\sim$ 8 digits. We briefly discuss possible use in quantum chemistry, including accuracy, algorithmic suitability and operating-system machine-implementation as an intrinsic function.
\end{abstract}


\maketitle 


%
%

%
\title{Direct evaluation of overlap integrals between Slater-type-orbitals}
\author{Michael J. Caola\fnref{caola@blueyonder.co.uk}}
\address{6 Normanton Rd, Bristol BS8 2TY, U.K.}
\section{Introduction}
Single-centre electronic atomic orbitals,
\begin{equation}
  \psi_{nlm} (\mathbf r ) = F_{nl}(r,\alpha)\;\mathcal{Y}_{ml}(\mathbf r),
\end{equation}
are building blocks in the quantum description of atoms, molecules, crystals and, hence in general of matter. In Eq(1) $F_{nl}(r)$ is a radial function, $\mathcal{Y}_{ml}=r^lY_{ml}$ is a $\it solid $ harmonic, and  $Y_{ml}$ the  familiar spherical (surface) harmonic. We state that  vital parts of Molecular Quantum Mechanics can be built with the overlap integral
\begin{eqnarray}
I({\mathbf R})=I({\mathbf R},n,l,m,n',l',m')=\nonumber\\
\int d{\mathbf r}\;\psi_{nlm} ({\mathbf  r})^{*} \;\psi_{n'l'm'}({\mathbf  {r-R}})
\end{eqnarray}
where vector ${\mathbf R}$ is the spatial  separation of the two orbital centres.
Important normalised  $F_{nl}(r)$ are the Gaussian-
\begin{equation}
F_{nl}(r)= .. e^{-\beta r^{2}}
\end{equation}
and Slater-type-orbitals (STOs)
\begin{eqnarray}
F_{nl}( r,\alpha)=F_{nl}(x,y,z,\alpha)=\frac{(2\alpha)^{n+1/2}}{\sqrt{(2n)!}} r^{n-1-l} e^{-\alpha r}, \nonumber\\
 r=r(x,y,z)=\sqrt{x^2+y^2+z^2}  \nonumber\\ 
\end{eqnarray}
where $\alpha$ is a screening constant. The STO is accepted as physically superior to the GTO, but numerical evaluation of its $I(\mathbf R)$ is more difficult; both have previously used special and associated functions, including: Fourier, Bessel, Laguerre, Gegenbauer, Gaunt, Hobson,  ..  . We shall next evaluate the  $I(\mathbf R)$ for STOs, as a direct single-stage integration, with no summations over 'special' functions.
 \section{Analysis}
From Eq(1) and Eq(2) we have
\begin{equation}
\psi_{n'l'm'}(\mathbf  {r-R})= F_{n'l'}(|\mathbf{ r-R}|,\alpha')\;\mathcal{Y}_{m'l'}(\mathbf{ r-R}),
\end{equation}
which is valid for arbitrary $F(r)$ and, with cartesian vectors $\mathbf r(x,y,z)$ and  $\mathbf R(X,Y,Z)$, will use
\begin{eqnarray}
 |\mathbf{ r-R}|= r'=r'(x,y,z)=\nonumber\\
\sqrt{(x-X)^2+(y-Y)^2+(z-Z)^2}.
\end{eqnarray}
Also, we use the $\it cartesian$ solid-harmonic$^{1,2}$ in Eq(5)
\begin{eqnarray}
\mathcal{Y}_{ml}(\mathbf r)= \mathcal{Y}_{ml}(x,y,z)=\left[\frac {(2l+1)(l+m)!(l-m)!}{4\pi}\right]^{1/2}   \nonumber\\  
\sum_{k=0}^{[(l-m)/2]}  \frac{(-x-iy)^{k+m}(x-iy)^{k}z^{l-m-2k}}{2^{2k+m}(k+m)!k!(l-m-2k)!},  \nonumber\\ 
l=0,1,2, ..\,\,\,\,\,;m=-l .. +l \nonumber\\
\end{eqnarray}
Thus with Eqs(5,6,7) in Eq(2) we have
\begin{center}
\begin{eqnarray}
I=I(X,Y,Z)= \nonumber\\
\int_{-\infty}^{\infty}dx\int_{-\infty}^{\infty}dy\int_{-\infty}^{\infty}\; dz\; F_{nl}(x,y,z,\alpha) \; \mathcal{Y}_{ml}(x,y,z) \nonumber\\
  F_{n'l'}(x-X,y-Y,z-Z,\alpha') \;  \mathcal{Y}_{m'l'}(x-X,y-Y,z-Z) \nonumber\\
\end{eqnarray} 
\end{center}
The $I(X,Y,Z)$ of Eq(8) can be evaluated by direct numerical computation and is valid for {\it arbitrary} orbitals specified by $ F_{nl}()$; this is our desired solution.
\par
For the case of an STO (8) becomes
\begin{eqnarray}
I=I(X,Y,Z)=\int_{-\infty}^{\infty} dx \int_{-\infty}^{\infty} dy \int_{-\infty}^{\infty} dz \nonumber\\
\frac{(2\alpha)^{n+1/2}}{\sqrt{(2n)!}} r^{n-1-l} e^{-\alpha r}  \mathcal{Y}_{ml}(x,y,z) \nonumber\\
\frac{(2\alpha')^{n'+1/2}}{\sqrt{(2n')!}} r'^{n'-1-l'} e^{-\alpha' r'}  \mathcal{Y}_{m'l'}(x-X,y-Y,z-Z), \nonumber\\
r=\sqrt{x^2+y^2+z^2} \;\;\;\; r'=\sqrt{(x-X)^2+(y-Y)^2+(z-Z)^2} \nonumber\\
\end{eqnarray}
\section{Numerical results}
We use Mathematica 10 and Maple 18 to calculate Eq(9). Each integral in the table below contains a comma (e.g. -0.117413789,53804531)  whose left 
figures agree with literature values$^{3-10}$: this is typically  8 digits. These data are collected in $^{9,10}$.
\begin{table*}
\begin{center}
\begin{tabular}{|c|c|c|c|c|c|c|c|c|c|c|c|c|} \hline
$n$&$l$&$m$&$n'$&$l'$&$m'$&$\alpha$&$\alpha'$&$R$&$\Theta$&$\Phi$&integral Eq(9)&\bf ref \\ \hline
                                                                                    \hline
1&0&0&2&1&0&10&2&1.4&0&0&-0.117413789,53804531&$^{3}$
\\ \hline
2&1&0&5&2&0&2&0.3&1.4&0&0&-0.23323008,22624455-2&$^{3}$  \\ \hline
3&2&0&3&2&0&7.5&2.5&5&$\pi$/3&2$\pi$/3&-0.68034002,4312253-4&$^{4} $\\ \hline
3&2&1&3&2&0&9.7&6.4&0.3&$\pi$/9&3$\pi$/4&0.013735076,44&$^{5}$  \\ \hline
8&0&0&8&0&0&5&1&1&0&0&0.0107437341,23348333&$^{6}$ \\ \hline
10&7&1&8&1&1&3&3&10&0&0&0.23447835,22183802-2&$^{7}$  \\ \hline
1&0&0&1&0&0&10&10&1.4&0&0&0.66799473,05543532-4&$^{8}$ \\ \hline
2&1&0&2&1&0&2&2&1.4&0&0&-0.10074038,66530121&$^{8}$  \\ \hline
\end{tabular}
\end{center}
\nopagebreak*
\end{table*}
\section{Discussion}
Our overlap integrals  for STOs Eq(9) agree with the literature to $\sim$ 8 digits
 We ask active experts (quantum chemists/physicists and computer-aware numerical-analysts) if our direct evaluation Eq(10) could be useful.
\par
Present methods, (sums over special function, SS) e.g. $^{3-10}$, to calculate Eq(9) are acceptable, so our proposed direct integration (DI) should consider $\it inter\, alia$:
\begin{itemize}
\item
What minimum accuracy is needed for quantum molecular calculations?  If $>$8 digits, then SS and DI give different values and we must ask
\item Which of SS and DI is more accurate (suitable)? It would be wrong to automatically assume that the established SS is more accurate: SS and DI are different methods needing expert  comparison. Along with accuracy we would like DI to have suitable and natural notation for its purpose, so we ask
\item
How would DI handle/evaluate any of the several integrals (of which the overlap is but one) occurring in quantum molecular mechanics? We sketch evaluation of coulomb $(ab
|cd)$, 'the two-electron, four centre integral, one of the greatest problems in quantum chemistry'$^{11}$: 
\end {itemize}
\begin{equation}
(ab|cd)=(12|34)=\int \frac{d1\,d2\,\psi_{a}(\bf r1) \psi_{b}(\bf r1)^{*} \psi_{c}(\bf r2) \psi_{d}(\bf r2)^{*} }{r_{12}},
\end{equation}
where
\begin{equation}
 \psi_{a}( r1)=\psi_{n_{a} l_{a}m_{a}}(\alpha_{a},x1,y1,z1), \,\,\,\,\,\,\,d1=dx1\,dy1\,dz1, \nonumber
\end{equation}
\begin{equation}
r_{12}=|\bf R+\bf r2-\bf r1| = \nonumber
\end{equation}
\begin{equation}
\sqrt{(X+x2-x1)^2+(Y+y2-y1)^2+(Z+z2-z1)^2},  \nonumber 
\end{equation}
etc., and is evaluated in Mathematica 10 in the same way used for overlap Eq(9).


\section{References}
\begin{thebibliography}{11}

\bibitem{1} Louck J S 1996 {\it Atomic, Molecular, $\&$ Optical handbook, ed Drake G W F} (New York: AIP Press) p~9
\bibitem{2} Caola M J 1978 {\it J. Phys. A: Math. Gen.} {\bf 11} L23-5
\bibitem{3} Barnett M P2003 {\it Int. J. Quantum Chem.} {\bf 95} 791-805
\bibitem{4} Ozdogan T 2004 {\it Collect. Czech Chem. Commun.}  {\bf 69} 279-91
\bibitem{5} Mamedov  B A 2004 {\it Chin. J. Phys.} {\bf 42} 176-81
\bibitem{6} Romanowski Z, Jailbout A F (2009) {\it  J. Math. Chem.} {\bf 45} 97-107
\bibitem{7} Yavuz M et al. 2005 {\it Commun. Theor. Phys.} {\bf 43} 151
\bibitem{8} Rico J F et al. 1988 {\it  J. Comput. Chem.} {\bf 9} 790
\bibitem{9} Ozdogan T, Nalcaci A 2012 {\it Int. J. Phys. Sciences} {\bf 7} 5378-90
\bibitem{10} Yukcu N 2012 {\it AWERProc. Inf. Tech. Comput. Sci.} {\bf 2} 61-6
\bibitem{11} Atkins P, Friedman R 2011 {\it Molecular Quantum Mechanics}(Oxford: O. U. P.)
\end{thebibliography}
\end{document}